# Comment on „An alternative approach for the determination of mean free paths of electron scattering in liquid water based on experimental data"

by Schild et al. doi.org/10.1021/acs.jpclett.9b02910 and arxiv.org/abs/1912.08469

Ruth Signorell

*Department of Chemistry and Applied Biosciences, Laboratory of Physical Chemistry,*
*ETH Zürich, Vladimir-Prelog-Weg 2, CH-8093, Zürich, Switzerland*
Email: rsignorell@ethz.ch

In a recent article [1], Schild et al. present what they call an alternative approach to the determination of mean free paths of electron scattering in liquid water. This by no means new approach is based on a very simplistic two channel model of electron scattering with only one fully elastic and one fully inelastic (total loss) channel. The parameterization consists of an elastic cross section, its angular dependence (« DCS »), and a cross section for electron loss by inelastic scattering. The two cross sections (or the equivalent MFPs) are determined from fits to experimental data (EALs from [2] and measured $\beta$-parameters from [3]). The elastic DCS is calculated ab initio for various cluster models. Furthermore, they claim to find elastic and inelastic mean free paths that are much shorter than those for amorphous ice from ref. [4]. We find that there are a number of issues both with the author's approach and with the comparison of their results with the literature. As outlined below, the reported values for elastic and inelastic mean free paths are questionable and conclusions regarding the difference between electron scattering in liquid water and amorphous ice are invalid.

**Fit to effective EAL values of [2]:** The EAL computed in [1] is not the same quantity as the effective EAL measured in [2]. Eq.(2) of [1] defines the EAL as the distance parameter of the exponential decay of the number of electrons detected as a function of the depth from where they originate. The effective EAL values of [2], by contrast, are derived from the total photoemission yield of the liquid. The absolute total yield is retrieved from the measured relative band intensities of liquid and vapor O(1s) bands using the vapor as an internal intensity standard. We also note that the derivation of the yield needs the experimental $\beta$-parameters from [3]. The yield is finally converted into an effective EAL under the assumption that all detected electrons originate from a shell with a thickness given by the value of the EAL. The definitions in [1] and [2] are inconsistent and as a result the MFPs derived in [1] are not correct.

**Ab initio cluster model of DCS :** The convergence of the cluster model towards liquid bulk is not demonstrated. The largest cluster they calculate is only the heptamer. Fig.S1 is misleading, as it mainly highlights the sharp elastic forward scattering peak. This peak does not significantly contribute to electron transport (elastic forward scattering changes neither kinetic energy nor direction). Relevant for electron transport and thus for MFP's is only the so-called momentum transfer cross section, which represents an effective isotropic elastic cross section. For this reason, scientist in the field usually quote the isotropic component of the elastic cross section (e.g. ref. [4]). Given that the EMFP and thus the total elastic cross section is a fit parameter and only the angular distribution is taken from the ab initio calculations in [1], the ab initio calculated elastic DCS become irrelevant for the determination of MFPs.

The cluster model seems even conceptually flawed. As described in [1], the calculations yield elastic cross sections for the overall scattering of an electron by a cluster as a whole in vacuo.

This does not say anything about the transport scattering inside the cluster, let alone in liquid water. On the one hand, the cluster model neglects the dielectric shielding in bulk water. On the other hand, even a very large cluster would still not yield transport scattering cross sections in liquid water. The ab initio calculations only describe the overall effect of scattering by a large cluster. To unravel the transport scattering inside the cluster would require a very involved partial wave analysis of the explicit wavepacket dynamics in a large cluster – far beyond the approach in [1].

**Comparison with literature results:** The comparison of EMFPs in Fig.2 of [1] with those of Michaud et al. [4] is completely flawed. The authors compare two different physical quantities. As pointed out above the EMFP of [1] includes the large contribution from elastic forward scattering, while [4] quotes only the isotropic component. No wonder that the values vastly differ. Therefore the authors' statement in the abstract that their EMFP is much shorter than that suggested by experiments for amorphous ice [4] is completely meaningless. The results of [1] do not say anything about how liquid water compares with amorphous ice. As already pointed out above, the comparison of IMFPs from [1] and [4] is not correct either.

Regarding the comments in the introductory part of [1] on the cluster results of [5] the authors appear not to have understood the physical reason why electron scattering cross sections in clusters necessarily differ from those determined for amorphous ice in [4]. This has nothing to do with a potential difference between liquid water and amorphous ice cross sections. The observed increase of scattering cross sections in clusters relative to bulk arises from the reduced dielectric shielding in the former.

**Uncertainties**: Throughout the comparison of their results with the literature the authors of [1] completely disregard any uncertainties (errors). On this basis they cannot possibly judge the agreement or disagreement between data sets (e.g. for ice and liquid MFPs). Just to give an example, according to their own sensitivity analysis of the fit (Fig.S3) the uncertainties quoted by Türmer et al. [3] for $\beta$ at 20eV would already translate into an uncertainty of the resulting IMFP of about a factor of two, while the IMFP at 10eV would only be determined to within an order of magnitude. Especially below 20 eV the experimental data sets [2,3] used for the fits in [1] are affected by particularly large uncertainties (partly arising from the background of secondary electrons).